\documentclass[showpacs,aps]{revtex4}
\usepackage{amsmath}
\usepackage{times}
\usepackage{epsfig}
\usepackage{amssymb}
\usepackage{color}

\usepackage{graphicx}
\usepackage{epsfig}
\usepackage{psfrag}
\newcommand{\dd}{\text{d}}

\newcommand{\ee}{\text{e}}

\newcommand{\p}{\partial}

\newcommand{\eps}{\varepsilon}

\begin{document}
\title{Equilibrium-like fluctuations in some boundary-driven open diffusive systems} 
\author{A. Imparato$^{(1)}$, V. Lecomte$^{(2)}$  and F. van Wijland$^{(3)}$}

\affiliation{$^{(1)}$Department of Physics and Astronomy, University of {A}arhus,
 Ny Munkegade, Building 1520, 8000 Aarhus, Denmark}

\affiliation{$^{(2)}$D\'epartement de Physique de la Mati\`ere Condens\'ee, Universit\'e de Gen\`eve, 24 quai Ernest Ansermet, 1211 Gen\`eve, Switzerland}

\affiliation{$^{(3)}$Laboratoire Mati\`ere et Syst\`emes Complexes 
(CNRS UMR 7057), Universit\'e Paris Diderot, 10 rue Alice Domon et L\'eonie Duquet, 75205 Paris cedex 13, France}

\begin{abstract}
There exist some boundary-driven open systems with diffusive dynamics whose particle current fluctuations exhibit universal features that belong to the Edwards-Wilkinson universality class. We achieve this result by establishing a mapping, for the system's fluctuations, to an equivalent open --yet equilibrium-- diffusive system. We discuss the possibility of observing dynamic phase transitions using the particle current as a control parameter.
\end{abstract}
\pacs{05.40.-a, 05.70.Ln} 
\maketitle

\section{Introduction}
In this work we consider systems of interacting particles undergoing diffusive dynamics, such as the  Symmetric Simple Exclusion Process (SSEP)~\cite{kipnisollavaradhan,spohn-1,liggett,kipnislandim}. Such systems, that can be described by the theory of fluctuating hydrodynamics, a coarse-grained description in terms of continuous degrees of freedom living in a continuous space, have already been the subject of intense investigation. For instance, Bertini {\it et al.}~\cite{bertinidesolegabriellijonalasiniolandim-1,bertinidesolegabriellijonalasiniolandim-2,bertinidesolegabriellijonalasiniolandim-3, bertinidesolegabriellijonalasiniolandim-5, bertinidesolegabriellijonalasiniolandim-6} have relied on fluctuating hydrodynamics to provide a quantitative analysis of large deviation properties of diffusive systems taken out of equilibrium by means of a boundary drive. Among the various large deviation properties investigated so far, those of the particle current play a special role. Indeed, current fluctuations have been known for a long time to be a central quantity since the work of Einstein~\cite{einstein,kubotodahashitsume} which established that in equilibrium the current variance is proportional to the diffusion constant; current fluctuations  characterize the likeliness of, and quantify, the system's excursions out of equilibrium. In the last decade, generic properties of these large deviation functions were discovered such as  the fluctuation theorem which determines how the large deviation function of the current is changed under time reversal~\cite{evanscohenmorriss,gallavotticohen,kurchan,lebowitzspohn,maesredigmoffaert,seifert,maes, gaspard, gallavotti2}. Parallel approaches~\cite{derridalebowitzspeer-1, derridalebowitzspeer-2,derridalebowitzspeer-3, bodineauderrida,bodineauderrida2, bodineauderrida3} which have been employed have revealed the possibility of new types of phase transitions where the particle current plays the role of a control parameter. It must also be mentioned that some results regarding the particle current statistics originate from exact solutions; this is the case for the totally asymmetric exclusion process~\cite{derridalebowitz} --a version of the SSEP where the particles' motion is strongly biased-- or for the SSEP~\cite{appertderridalecomtevanwijland}, with periodic boundary conditions.\\

Fluctuating hydrodynamics not only encompasses interacting particle systems, but also applies to models involving at the microscopic level already continuous degrees of freedom, such as the Kipnis-Marchioro-Presutti~\cite{kipnismarchioropresutti} (KMP) model of interacting harmonic oscillators that has served as a testbench to investigate the statistical properties of heat conduction. Both the SSEP and the KMP models will be the main focus of our efforts in the sequel, though we shall strive to keep our discussion general when possible.\\

Our central motivation is  to investigate the role of finite-size effects in one-dimensional open driven diffusive systems. We have been inspired by the  results of Appert {\it et al.}~\cite{appertderridalecomtevanwijland} and by those of Derrida and Lebowitz~\cite{derridalebowitz}. In the former universal properties were seen to emerge in the statistics of the particle current for an equilibrium diffusive system with periodic boundary conditions, with the possibility, for certain classes of diffusive systems, to exhibit a current-driven dynamic phase transition. In the latter, where mutually excluding particles are subjected to a bulk electric field that drives the system far from equilibrium, universal features belonging to a different universality class have also been observed. Besides, Bodineau and Derrida~\cite{bodineauderrida2, bodineauderrida3} have shown that for a weakly bulk and boundary driven SSEP, a dynamic phase transition takes place.\\

In the present work our interest goes to open systems, maintained out of equilibrium by putting them in contact with particle reservoirs at unequal chemical potentials, but the bulk dynamics itself remains reversible. Thus the nonequilibrium nature of our systems does not arise from an external bulk field but only from a boundary drive. We ask the following questions: (i) Do universal features in the particle current appear in an open system? If so, do they depend on the system being possibly driven out of equilibrium by a chemical potential gradient? (ii) Is the existence of a particle current capable of inducing a dynamic phase transition?\\

Before entering the technicalities of our work, we would like to phrase the answers we have come up to issues (i) and (ii). To question (i) we have the partial answer that at least for a class of systems --to which the SSEP and the KMP model belong-- the current distribution does indeed display universal features. The latter do not depend on the system being in or out of equilibrium, and, quite remarkably, they are the same for an open system as for a closed system~\cite{appertderridalecomtevanwijland}. They belong to the Edwards-Wilkinson universality class~\cite{edwardswilkinson}. To question (ii) the answer is not straightforward: we have found at least a family of physical systems (among which the SSEP) in which the current large deviation function displays some singularity, indicating the existence of a dynamic phase transition depending on the scaling of the current one forces through.\\

We shall begin in section \ref{sectionrecall} by recalling what is known on the statistics of the current in an open boundary driven diffusive system. Section \ref{sectionfluctuations} is devoted to careful analysis of finite-size effects leading to establishing that in some cases fluctuations exhibit universal features. This section is supplemented by an appendix that describes the cases of the SSEP and the KMP model in detail. Our conclusions and yet open problems are gathered in section \ref{sectionoutlook}.

\section{Current large deviations in diffusive in boundary-driven open systems}\label{sectionrecall}
We consider a one-dimensional lattice with $L$ sites, whose state at time $t$ is characterized by the local numbers $n_j(t)$'s (which may be discrete or continuous variables), $j=1,\ldots, L$. Our starting point is the assumption that, in the large $t$ and $L$ limit, with $t/L^2$ fixed, there exists a Langevin equation for a density field $\rho(x,\tau)=n_j(t')$, with $x=j/L$ and $\tau=t'/L^2$ defined over $x\in[0,1]$ and $\tau\in [0,t/L^2]$ which evolves according to
\begin{equation}\label{wkb1}
\p_\tau\rho=\p_x(D(\rho(x,\tau))\p_x\rho(x,\tau)-\xi(x,\tau))
\end{equation}
where the Gaussian white noise $\xi$ has correlations $\langle\xi(x,\tau)\xi(x',\tau')\rangle=\frac{\sigma(\rho(x,\tau))}{L}\delta(x-x')\delta(\tau-\tau')$ that decay to zero as the inverse system size. The phenomenological coefficients $D(\rho)$ (the diffusion constant) and $\sigma(\rho)$ depend on some of the details of the underlying microscopic dynamics. The system size $L$ and the observation time $t$ are large with respect to microscopic space and time scales. The system is in contact at both ends with reservoirs that fix the value of $\rho$ at all times to be $\rho_0$ at $x=0$ and $\rho_1$ at $x=1$.\\

Our interest lies in the statistics of the total particle current $Q(t)$ accumulated up until time $t$ and its large deviation properties, which, in terms of the field $\rho$, is formally expressed as
\begin{equation}
Q(t)=L^2\int_0^1\dd x\int_0^{t/L^2}\dd\tau \left(-D(\rho(x,\tau))\p_x\rho(x,\tau)+\xi(x,\tau)\right)
\end{equation}
Our purpose is to determine 
\begin{equation}
\pi(j)=\lim_{t\to\infty}\frac{\ln\text{Prob}\{Q(t)=j\; t\}}{t}
\end{equation}
or, equivalently, its Legendre transform $\psi(s)=\text{max}_j\{\pi(j)-sj\}$ that can be obtained from the generating function of $Q$ as follows
\begin{equation}\label{defpsi}
\psi(s)=\lim_{t\to\infty}\frac{\ln \langle \ee^{-s Q(t)}\rangle}{t}
\end{equation}
Using the Janssen-De Dominicis formalism~\cite{janssen}, we see that the generating function $\langle \ee^{-s Q(t)}\rangle$ can be cast in the form of a path-integral over two fields,
\begin{equation}\label{wkb2}
\langle \ee^{-s Q(t)}\rangle=\int\mathcal{D}\bar{\rho}\mathcal D\rho\;
\ee^{-L S[\bar{\rho},\rho]}
\end{equation}
where the action $S$ is given by
\begin{equation}\label{action}
S[\bar{\rho},\rho]=\int_0^1\dd x\int_0^{t/L^2}\dd\tau\left[\bar{\rho}\p_\tau\rho+D(\rho)\p_x\bar{\rho}\p_x\rho-\frac{\sigma}{2}(\p_x\bar{\rho}-\lambda)^2-\lambda D\p_x\rho\right]
\end{equation}
where $\lambda=sL$, and the path-integral runs over functions verifying the boundary conditions $\rho(0,\tau)=\rho_0$, $\rho(1,\tau)=\rho_1$, and $\bar{\rho}(0,\tau)=\bar{\rho}(1,\tau)=0$. It is very clear from the expression of the noise in (\ref{wkb1}) or from the path-integral (\ref{wkb2}) that a semi-classical-like expansion is valid in the weak-noise limit, which, translated in our language, is synonymous for a large system-size expansion. In short, the path-integral (\ref{wkb2}) is dominated by the saddle point of $S$. The reader is referred to Kurchan's lectures~\cite{kurchan2} for a pedagogical account exploiting this language. We first change the response field into $\tilde{\rho}(x,\tau)=\bar{\rho}(x,\tau)-\lambda x$. We write the saddle-point equations and we assume the saddle is reached for time-independent profiles $\tilde{\rho}_c(x),\rho_c(x)$. Sufficient conditions under which this is so have been discussed by Bertini {\it et al.}~\cite{bertinidesolegabriellijonalasiniolandim-6}. Assuming this is indeed the case, the saddle point equations $\frac{\delta S}{\delta \rho}=0$, $\frac{\delta S}{\delta \tilde{\rho}}=0$ read
\begin{equation}
\p_\tau\rho=\p_x(D\p_x\rho)-\p_x(\sigma\p_x\tilde{\rho}),\;\; -\p_\tau\tilde{\rho}=\p_x(D\p_x\tilde{\rho})+\frac{\sigma'}{2}(\p_x\tilde{\rho})^2
\end{equation}
which, assuming a stationary solution, lead to
\begin{equation}\label{saddle1}
D^2(\rho_c)(\p_x\rho_c)^2=K_1^2+K_2\sigma(\rho_c),\;\;\p_x{\tilde{\rho}_c}=\frac{D(\rho_c)\p_x\rho_c+K_1}{\sigma(\rho_c)}
\end{equation}
where $K_1$ and $K_2$ are $\lambda$-dependent constants. With these equations, one may verify that the action evaluated at the saddle reads $S[\tilde{\rho}_c,\rho_c]=(t/L^2)K_2/2$. We shall denote by $\mu(\lambda)=-K_2/2$ (our definition of $\mu$  differs from that of \cite{derridadoucotroche,bodineauderrida} by a factor $1/L$). With these notations, for the large deviation function introduced in (\ref{defpsi}), we thus have $\psi(s)=\frac{\mu(sL)}{L}$. In practice, to determine $\mu(\lambda)$ explicitly, one must solve the differential equations (\ref{saddle1}) and fix the constants $K_1$ and $K_2$ by means of the appropriate boundary conditions. A few comments are in order: these results are not new and they were first derived by Bodineau and Derrida~\cite{bodineauderrida}. We propose as an illustration the explicit expression for $\mu(\lambda)$ when the diffusion constant $D$ is independent of the local density (we take $D=1$) and when the noise strength $\sigma(\rho)$ is a simple quadratic function. For $\sigma(\rho)=c_2\rho^2+c_1\rho$ we find that (see appendix A)
\begin{equation}\label{muomega}
\mu(\lambda)=\left\{
\begin{array}{ll}
-\frac{2}{c_2}(\operatorname{arcsinh}\sqrt{\omega})^2&\text{ for }\omega>0\\
+\frac{2}{c_2}(\arcsin\sqrt{-\omega})^2&\text{ for }\omega<0
\end{array}
\right.
\end{equation}
where $\omega(\lambda,\rho_0,\rho_1)$ is the auxiliary variable given by
\begin{equation}\label{def-omega-general}
\omega(\lambda,\rho_0,\rho_1)=\frac{c_2}{c_1^2}(1-\ee^{c_1\lambda/2})\left(c_1(\rho_1-\ee^{-c_1\lambda/2}\rho_0)-c_2(\ee^{-c_1\lambda/2}-1)\rho_0\rho_1\right)
\end{equation}
For the SSEP, $\sigma(\rho)=2\rho(1-\rho)$ and one recovers the known~\cite{derridadoucotroche,jordansukhorukovpilgram} result (the notation $z=\ee^{-\lambda}$ is used in the formula (2.14) of~\cite{derridadoucotroche}), namely
\begin{equation}\label{def-omega-SSEP}
\omega(\lambda,\rho_0,\rho_1)=(1-\ee^\lambda)(\ee^{-\lambda}\rho_0-\rho_1-(\ee^{-\lambda}-1)\rho_0\rho_1)
\end{equation}
Another solvable model is the KMP chain of coupled harmonic oscillators, for which $D=1$ and $\sigma(\rho)=4\rho^2$ (for KMP, $\rho$ stands for the local potential energy field), for which we also have (\ref{muomega}) but where the variable $\omega$ is now given by
\begin{equation}\label{def-omega-KMP}
\omega(\lambda,\rho_0,\rho_1)=\lambda(2(\rho_0-\rho_1)-4\lambda\rho_0\rho_1)
\end{equation}
The case $c_2>0$ and $c_2<0$ are qualitatively different. In the latter, $\mu(\lambda)$ is defined over the whole real axis and is unbounded from above, while in the former $\mu(\lambda)$ is defined over a finite interval of $\lambda$ whose ends correspond to infinite currents produced by the build-up of infinite densities. For example, with $c_2=4$ and $c_1=0$, that is for the KMP model, $-\frac{1}{2\rho_1}<\lambda<\frac{1}{2\rho_0}$.\\

Finally, it is important to realize that $\mu(\lambda)$ is the leading order term in a large system-size series expansion. The origin of finite-size corrections is twofold. Of course there will be finite-size corrections arising from integrating out the modes describing fluctuations around the optimal profile $\{\tilde{\rho}_c,\rho_c\}$. However, fluctuating hydrodynamics, by definition, is unable to capture the details of the microscopic systems it describes. It must therefore be expected that model-dependent finite-size corrections will also emerge. We now proceed with determining the finite-size contribution of fluctuations around the saddle of the action (\ref{action}) within the framework of fluctuating hydrodynamics (that is, temporarily omitting contributions arising from the underlying discreteness of the lattice).

\section{Fluctuations and universal behavior}\label{sectionfluctuations}
\subsection{Evaluating a determinant}
As in any saddle point calculation, we obtain finite size-corrections to the leading order result $\langle \ee^{-s Q}\rangle\simeq \ee^{\mu(s L) t}$ by introducing, in the path-integral (\ref{wkb2}), the fluctuations around the optimal profile $\tilde{\rho}_c$ and $\rho_c$: $\phi(x,\tau)=\rho(x,\tau)-\rho_c(x)$ and $\bar{\phi}(x,\tau)=\tilde{\rho}(x,\tau)-\tilde{\rho}_c(x)$. Then we expand the action (\ref{action}) to quadratic order in $\phi$ and $\bar{\phi}$:
\begin{equation}\label{action-fluct}\begin{split}
S=-\frac{\mu(\lambda) t}{L^2}+\int\dd x\dd\tau&\left(
\bar{\phi}\p_\tau\phi+D\p_x\bar{\phi}\p_x\phi  +D'\p_x\tilde{\rho}_c\phi \p_x\phi+D'\p_x\rho_c\p_x\bar{\phi}\phi  +\frac{D''}{2}\p_x\tilde{\rho}\p_x\rho\phi^2  \right.\\  
&\left.-\frac{\sigma}{2}(\p_x\bar{\phi})^2-\sigma'\p_x\tilde{\rho}_c\phi\p_x\bar{\phi}-\frac{\sigma''}{4}(\p_x\tilde{\rho}_c)^2\phi^2\right)
\end{split}\end{equation}
where $D$, $\sigma$, and their derivatives with respect to the density, are evaluated at $\rho_c(x)$. The goal is to integrate out the quadratic action (\ref{action-fluct}) with respect to the fields $\bar{\phi}$ and $\phi$. This is the procedure that was followed in \cite{appertderridalecomtevanwijland} and that we carry out here as well. However, unlike the case of periodic boundary conditions dealt with in \cite{appertderridalecomtevanwijland}, in the present case, the quadratic action is not readily diagonalizable for its coefficients are space-dependent. It so happens that for one particular family of models, those for which $D(\rho)$ is constant and $\sigma(\rho)$ is quadratic in $\rho$, this can actually be achieved. This remains a nontrivial task, given that the quadratic form to diagonalize in (\ref{action-fluct}) still possesses space-dependent coefficients. We have not been able deal with arbitrary $D$ and $\sigma$.
\subsection{Constant $D$ and quadratic $\sigma$}
We specialize the action (\ref{action-fluct}) to a constant $D$ and a quadratic $\sigma$. After performing the change of fields 
\begin{equation}\label{cov}
\phi=(\p_x\tilde{\rho}_c)^{-1}\psi+\p_x\rho_c\bar{\psi},\;\;\bar{\phi}=\p_x\tilde{\rho}_c\bar{\psi}
\end{equation}
we note that (\ref{action-fluct}), after tedious rearrangements, becomes
\begin{equation}\label{action-fluct-2}\begin{split}
S=-\frac{\mu(\lambda)t}{L^2}+\int\dd x\dd\tau\left(\bar{\psi}\p_\tau\psi+D\p_x\bar{\psi}\p_x\psi -\mu(\lambda)(\p_x\bar{\psi})^2-\frac{\sigma''}{4}\psi^2\right) 
\end{split}\end{equation}
It is remarkable that (\ref{action-fluct-2}) is now a quadratic form that can be diagonalized with standard stationary waves $\{\sin qx\}_q$ with Fourier modes indexed by $q=\pi n$, $n\in \mathbb{N}^*$. By comparison to (\ref{action-fluct}), we can interpret (\ref{action-fluct-2}) as being the action corresponding to an {\it equilibrium} open system whose current fluctuations we study as a function of the conjugate variable $\frac{\mu(sL)}{L}$. Performing the change of variables (\ref{cov}) has allowed us to map the fluctuations onto those of an open system in contact with two reservoirs at equal densities.\\

After integrating out the $\psi$ and $\bar{\psi}$ fields, one arrives at 
\begin{equation}\label{resultat1}
\psi_{\text{\tiny FH}}(s)=\frac{1}{L}\mu(sL)+\frac{D}{8L^2}{\mathcal F}\left(\frac{\sigma''}{2D^2}\mu(s L)\right)
\end{equation}
where the FH index stands for ``Fluctuating Hydrodynamics'' where the function $\mathcal F$ has the expression
\begin{equation}\label{universalF}
{\mathcal F}(u)=-4\sum_{q=n\pi,\, n\geq 1}\left(q\sqrt{q^2-2u}-q^2+u\right)
\end{equation}
Equation (\ref{resultat1}) is the first new result of this work. It indicates that for systems whose fluctuating hydrodynamics description relies on a constant diffusion constant $D$ and a quadratic noise strength $\sigma(\rho)$, current fluctuations involve a universal scaling function $\mathcal F$. It is remarkable that exactly the same function $\mathcal F$ has appeared in the study of current fluctuations in closed systems {\it in equilibrium}, with a different scaling variable though. As can be seen from its explicit expression (\ref{universalF}), the scaling function $\mathcal F$ has a singularity when its argument approaches $\pi^2/2$ from the right real axis. In \cite{appertderridalecomtevanwijland, bodineauderridalecomtevanwijland} this was interpreted as the presence of a first-order dynamic phase transition, for systems with periodic boundary conditions. In the present case, this also opens up the possibility of a similar phase transition on condition that there exists a regime of $\lambda$ for which
\begin{equation}\label{cond-pt}
\frac{\sigma''}{2D^2}\mu(\lambda)>\frac{\pi^2}{2}
\end{equation}
Before we discuss whether a phase transition can indeed occur, we must address another pending issue.

\subsection{Microscopic details matter}
The expression for $\psi_\text{\tiny FH}(\lambda/L)=\frac{1}{L}\mu(\lambda)+\frac{D}{8L^2}{\mathcal F}\left(\frac{\sigma''}{2D^2}\mu(\lambda)\right)$ obtained from fluctuating hydrodynamics ignores the possibility that finite-size corrections of the same $\mathcal{O}(L^{-2})$ order as the universal corrections will appear when one relies on the original model defined on a lattice. In the appendix, which is based upon methods developed by Tailleur {\it et al.}~\cite{tailleurkurchanlecomte-1,tailleurkurchanlecomte-2}, we are able to evaluate the contribution of lattice effects for two specific models. We show that for the SSEP and for the KMP model they do introduce ${\mathcal O} (L^{-2})$ terms that add up to the universal contribution found from fluctuating hydrodynamics. Let us look into those microscopic details more precisely, first for the SSEP, then for the KMP model.\\

The open and driven SSEP consist of particles hopping to either of their nearest neighbor sites on a lattice of $L$ sites, in contact with particle reservoirs connected to sites $1$ and $L$. Particles are injected into site $1$ (resp. $L$) with a rate $\alpha$ (resp. $\delta$) and are removed from site $1$ (resp. $L$.) with rate $\gamma$ (resp. $\beta$). These reservoirs impose densities $\rho_0=\frac{\alpha}{\alpha+\gamma}$ and $\rho_1=\frac{\delta}{\beta+\delta}$ at sites $1$ and $L$. While in the fluctuating hydrodynamic formulation the reservoirs enter current statistics through $\rho_0$ and $\rho_1$ only, when one wishes to capture phenomena beyond leading order, lattice effects and microscopic details start playing a role. For the SSEP, as presented in  appendix B, introducing the auxiliary constants $a=\frac{1}{\alpha+\gamma}$ and $b=\frac{1}{\beta+\delta}$, we find that
\begin{equation}\label{SSEP-lat}\begin{split}
\psi(s)=&\psi_{\text{\tiny FH}}(s)-\frac{a+b-1}{L^2}\mu(\lambda)+{\mathcal O}(L^{-3})
\\=&\frac{1}{L}\mu(\lambda)-\frac{a+b-1}{L^2}\mu(\lambda)+\frac{D}{8L^2}{\mathcal F}\left(\frac{\sigma''}{2D^2}\mu(\lambda)\right)+{\mathcal O}(L^{-3})
\end{split}\end{equation}
Note that this result is compatible with the exact expressions of the first three cumulants of the current obtained in~\cite{derridadoucotroche}.

The KMP model is also a lattice model in which $L$ harmonic oscillators whose positions $x_j$ are coupled (we use the It\^o convention and the Giardin\`a {\it et al.}~\cite{giardinakurchanredig} version of the KMP model)
\begin{equation}\label{defLangevinKMP}
2\leq j\leq N-1,\;\;\frac{\dd x_j}{\dd t}=-x_j+x_{j+1}\eta_{j,j+1}- x_{j-1}\eta_{j-1,j}
\end{equation}
and the chain is in contact at both ends with heat baths imposing temperatures $T_1$ and $T_L$, 
\begin{equation}
\frac{\dd x_1}{\dd t}=-\left(\gamma_1 +\frac 12\right)x_1-\sqrt{2\gamma_1 T_1}\xi_1+ x_2\eta_{1,2},\;\frac{\dd x_L}{\dd t}=-\left(\frac 12 +\gamma_L \right)x_L+\sqrt{2\gamma_L T_L}\xi_L- x_{L-1}\eta_{L-1,L}
\end{equation}
where $\xi_1$, $\xi_L$, $\eta_{j,j+1}$ (for $1\leq j\leq L-1$) are Gaussian white noises with variance unity, and $\gamma_1$, $\gamma_L$ set the time-scale of the energy exchange with each reservoir. We refer the reader to Giardin\`a {\it et al.}~\cite{giardinakurchanredig,giardinakurchanredigvafayi} for further details and connections between the SSEP and KMP. It is also shown in appendix B that for the KMP model we have
\begin{equation}\label{KMP-lat}
\psi(s)=\frac{1}{L}\mu(\lambda)-\frac{\frac{1}{2\gamma_1}+\frac{1}{2\gamma_L}-1}{L^2}\mu(\lambda)+\frac{D}{8L^2}{\mathcal F}\left(\frac{\sigma''}{2D^2}\mu(\lambda)\right)+{\mathcal O}(L^{-3})
\end{equation}
Both (\ref{SSEP-lat}) and (\ref{KMP-lat}) reveal that taking into account microscopic details of the systems leads, as expected, to nonuniversal corrections to the current large deviation function. Whatever the form of these nonuniversal contributions to $\psi(s)$, it can be seen that the relevant piece of information regarding the possibility of a phase transition is contained in the universal part of $\psi_{\text{\tiny FH}}(s)$.

\subsection{Is a dynamic phase transition possible?}
For systems having a constant diffusion constant $D$ (which we set to $D=1$) and a quadratic $\sigma(\rho)=c_2\rho^2+c_1 \rho$, the explicit expression of $\mu(\lambda)$ obtained in (\ref{muomega}) allows us to probe the criterion (\ref{cond-pt}) for the existence of a phase transition. Working at fixed $\lambda=s L$, no first-order phase transition can occur because the condition (\ref{cond-pt}), or equivalently, $\omega(\lambda,\rho_0,\rho_1)=1$, cannot be fulfilled on the real axis of $\lambda$. In the original variable $s$, however, things are different, since in the large system size limit, and for $c_2<0$ only, the singularity in the complex plane of $s$ eventually hits the real axis at $s=0$. To be more explicit, using (\ref{def-omega-general}), one notices that for $c_2<0$
\begin{equation}
\lambda\to\infty,\;\;\mu(\lambda)\simeq - \frac{c_1^2}{2 c_2 }  \lambda^2
\end{equation}
so that, after inserting into (\ref{muomega}) and taking the asymptotics, one arrives at
\begin{equation}\label{thirdo}
\lim_{L\to\infty}\frac{\psi_Q(s)}{L}=\frac{2 c_1^2}{c_2}s^2+\frac{c_1^3}{3\pi}|s|^3+o(s^3)
\end{equation}
The singularity at $s=0$ reflects the existence of a dynamic transition in terms of the total particle current, but of higher order. The same transition existed for systems with periodic boundary conditions (see (62) of \cite{appertderridalecomtevanwijland}) and was noted earlier by Lebowitz and Spohn (in (A.12) of \cite{lebowitzspohn}). The effects of this transition can be seen on the correlation functions~\cite{bodineauderridalecomtevanwijland} which become long-ranged. Note also that in this scaling limit (\ref{thirdo}) does not depend on the reservoir densities anymore, because the optimal profile able to carry such large currents settles to density $\frac 12$, but in vanishingly small region around the system's boundaries.\\

For systems with $c_2>0$, that is systems with attractive interactions, such as the KMP model (for which  $\sigma(\rho)=4\rho^2$), no phase transition can be observed, but the trivial one occurring at infinite densities (akin to a Bose condensation). There exists a set of numerical simulations by Hurtado and Garrido~\cite{hurtadogarrido} for the KMP model which actually confirm that no phase transition is observed. This negative result is in contrast with --but does not contradict-- that of Bertini {\it et al.}~\cite{bertinidesolegabriellijonalasiniolandim-6, appertderridalecomtevanwijland} in which it was shown that a phase transition exists, for periodic boundary conditions, when $\sigma''>0$.

\section{Outlook}\label{sectionoutlook}
We have shown that in a family of diffusive systems driven out of equilibrium by a chemical potential gradient, the total particle current exhibits universal fluctuations. These belong to the Edwards-Wilkinson universality class and they are of the same form as that previously found in closed equilibrium systems. Our results apply to diffusive systems characterized by a constant $D$ and a quadratic $\sigma$. We have used a mapping of the system's fluctuations to those of an equivalent open system in equilibrium. We have hints that this mapping can be extended beyond quadratic fluctuations: for the SSEP, we can actually prove that a similar mapping applies to the full process~\cite{lecomteetal}. Our main concern lies in that our results are indeed limited to the case $D$ constant and $\sigma$ quadratic. It would be of great interest to find out whether similar universal properties hold for generic $D$ and $\sigma$. Perhaps is this a fortuitous coincidence, but Tailleur {\it et al.}~\cite{tailleurkurchanlecomte-1, tailleurkurchanlecomte-2} ran into similar restrictions when mapping the density profile large deviations in boundary-driven diffusive systems onto their equilibrium counterparts. We see here subjects for future research.\\

\noindent {\bf Acknowledgment} We thank Bernard Derrida and Thierry Bodineau for countless
discussions and critical comments. VL was supported in part by the
Swiss NSF under MaNEP and Division II. AI gratefully acknowledges the hospitality and support of Laboratoire Mati\`ere et Syst\`emes Complexes-Universit\'e Paris Diderot while this work was done.

\section*{Appendix A}
In this appendix we prove (\ref{muomega},\ref{def-omega-general}). We assume that 
\begin{equation}
  D(\rho)= 1 \quad;\qquad 
  \sigma(\rho)=c_2\rho^2+c_1\rho
\end{equation}
with $c_1>0$ and the boundary conditions $\rho(0)=\rho_0$, $\rho(1)=\rho_1$. In order to find the explicit expression of $\mu(\lambda)$ we start from the implicit equation found in \cite{bodineauderrida} specialized to $D=1$, namely
\begin{equation}
   \mu(\lambda)= 
   -K \left[ \int_{\rho_0}^{\rho_1} \dd\rho
   \frac 1 {\sqrt{1+2K\sigma}}\right]^2
  \label{eq:mu_of_lambda}
\end{equation}
with $K$ determined by
\begin{equation}
  \lambda = 
  \int_{\rho_0}^{\rho_1} \dd\rho\frac 1 \sigma \Bigg[\frac 1 {\sqrt{1+2K\sigma}} -1\Bigg]
  \label{eq:lambda_K}
\end{equation}
We know that the optimal profile  verifies
\begin{equation}
  \partial_x \rho = 
  q \sqrt{1+2K\sigma}
\end{equation}
the solution of which takes the form
\begin{equation}
  \rho(x) = -\frac{c_1}{c_2}+ f
   \sinh\big[2(\theta_0+(\theta_1-\theta_0)x)\big]
  \label{eq:saddle_rho}
\end{equation}
provided $f$, $\theta_0$ and $\theta_1$ verify
\begin{align}
 (\theta_1-\theta_0)^2 &= \frac 12 c_2  K q^2
\\
 f^2 &= \frac{2c_2-c_1^2 K}{4 c_2^2 K}
 \label{eq:link_K_f}
\end{align}
and the boundary conditions $\rho(0)=\rho_0$, $\rho(1)=\rho_1$.
One performs the change of variable
\begin{equation}
  \dd x = \frac 1 q \frac{\dd\rho}{\sqrt{1+2K\sigma}}
  \label{eq:x_to_rho}
\end{equation}
in (\ref{eq:mu_of_lambda}) and (\ref{eq:lambda_K}). This yields
\begin{align}
  \mu(\lambda) &= -\frac 2{c_2} (\theta_1-\theta_0)^2
 \label{eq:muthetaab}
\\
  \lambda &=+
  \frac {2}{c_1} \ln \frac
   {c_1\rho_0\cosh 2\theta_1\:-\:\sqrt{c_1^2+4c_2^2f^2}\:\rho_0\sinh 2\theta_1}
   {c_1\rho_1\cosh 2\theta_0\:-\:\sqrt{c_1^2+4c_2^2f^2}\:\rho_1\sinh 2\theta_0}
 \label{eq:lambdaKtheta}
\end{align}
where we have used (\ref{eq:link_K_f}) to eliminate $K$ in favor of $f$, together with the boundary conditions
\begin{equation}
  \rho_0=-\frac{c_1}{2c_2}+f\sinh 2\theta_0\ ,\qquad \rho_1=-\frac{c_1}{2c_2}+f \sinh 2\theta_1
  \label{eq:tatbrhoarhob}
\end{equation}
We are left with eliminating $f$, $\theta_0$ and $\theta_1$ from (\ref{eq:muthetaab}-\ref{eq:tatbrhoarhob}).
Isolating first $f^2$ from (\ref{eq:lambdaKtheta}) one obtains
\begin{equation}
  f^2 \ = \
  \frac{c_1^2}{4c_2^2}\:
  \frac
   {z^2 \rho_0^2 - 2 z \rho_0\rho_1 \cosh 2(\theta_1-\theta_0) +\rho_1^2}
   {\big(z\,\rho_0\sinh 2\theta_1 - \rho_1\sinh 2\theta_0  \big)^2}
 \label{eq:last_step}
\end{equation}
where we have set $z=\ee^{-c_1\lambda/2}$.  Grouping $f^2$ with the
square at the denominator in (\ref{eq:last_step}), one eliminates $f$
using the boundary conditions (\ref{eq:tatbrhoarhob}). This enables us to
isolate $\cosh 2(\theta_1-\theta_0)$ in (\ref{eq:last_step}) and one gets
$\sinh^2(\theta_1-\theta_0)=\omega$ with
\begin{equation}
\omega=
 \frac{c_2}{c_1^2 } \big(1-z^{-1}\big)
 \big( c_1(\rho_1-z\rho_0) -c_2(z-1)\rho_0\rho_1 \big)
\end{equation}
and finally
\begin{equation}
  \mu(\lambda) = 
  \begin{cases}
    -\frac{2}{c_2}\big(\!\operatorname{argsinh}\sqrt{\omega}\big)^2 
    & \text{for $\omega>0$}
\\
    +\frac{2}{c_2}\big(\!\operatorname{arcsin}\sqrt{-\omega}\big)^2 
    & \text{for $\omega<0$}
  \end{cases}
\end{equation}

In the limit $c_1\to 0$ which is relevant for KMP, $\omega$ becomes
\begin{equation}
  \omega = 
  \frac 14 c_2 \lambda \big( 2(\rho_0-\rho_1)-c_2 \lambda\rho_0\rho_1 \big)
\end{equation}

\section*{Appendix B}
\subsection{Simple Symmetric Exclusion Process}
We consider a SSEP on a one-dimensional lattice with $L$ sites, in which particles are injected to the leftmost site $j=1$ (resp. rightmost site $j=L$) with rate $\alpha$ (resp. $\delta$), and removed with rate $\gamma$ (resp. $\beta$). The master operator governing the evolution of a microscopic configuration of occupation numbers $\{n_j\}_{j=1,\ldots,L}$ can be written in the form
\begin{equation}\label{evolQtot}\begin{split}
\mathbb{W}(s)=&
\sum_{1\leq k\leq L-1}\left(\ee^{s}\sigma_k^+ \sigma_{k+1}^-+\ee^{-s}\sigma_k^- \sigma_{k+1}^+-\hat{n}_k (1-\hat{n}_{k+1})-\hat{n}_{k+1}(1-\hat{n}_k)\right)\\
&+\alpha(\ee^{-s} \sigma_1^+-(1-\hat{n}_1))
+\gamma\left(\ee^{s}\sigma_1^--\hat{n}_1\right)
+\delta(\ee^{-s}\sigma_L^+-(1-\hat{n}_L))
+\beta\left(\ee^{s}\sigma_L^--\hat{n}_L\right)
\end{split}\end{equation}
In (\ref{evolQtot}) we are using a spin basis: the eigenvalue of the Pauli matrix $\sigma_j^z$ is 1 if site $j$ is occupied, and $-1$ if it is empty ($\hat{n}_j=\frac{1+\sigma_j^z}{2}$ has the eigenvalue $n_j=0$ or $1$).
We now remark that 
\begin{equation}\label{berry}
\ee^{s\sum_{j=1}^Lj\hat{n}_j}\mathbb{W}(\lambda)\ee^{-s\sum_{j=1}^Lj\hat{n}_j}=\mathbb{W}_L(s(L+1))
\end{equation}
where $\mathbb{W}_L$ is the operator counting the total current across site $L$ only, the expression of which reads
\begin{equation}\label{evolQ1}\begin{split}
\mathbb{W}_{L}(s')=&
\sum_{1\leq k\leq L-1}\left(\sigma_k^+ \sigma_{k+1}^-+\sigma_k^- \sigma_{k+1}^+-\hat{n}_k (1-\hat{n}_{k+1})-\hat{n}_{k+1}(1-\hat{n}_k)\right)\\
&+\alpha( \sigma_1^+-(1-\hat{n}_1))
+\gamma\left(\sigma_1^--\hat{n}_1\right)
+\delta(\ee^{s'}\sigma_L^+-(1-\hat{n}_L))
+\beta\left(\ee^{-s'}\sigma_L^--\hat{n}_L\right)
\end{split}\end{equation}
with $s'$ being conjugate to the time-integrated current through site $i=L$. Owing to (\ref{berry}), The largest eigenvalue of $\mathbb{W}_L(s(L+1))$ is $\psi(s)$, that is the largest eigenvalue of (\ref{evolQtot}). We use, for each lattice site, a Holstein-Primakoff like representation~\cite{tailleurkurchanlecomte-2}
\begin{equation}
\sigma^+=1-F+F^+-2 F F^++F^2 F^+,\; \sigma^-=F-F^2 F^+
\end{equation}
which also leads to $\hat{n}=F+F F^+-F^2 F^+$. The bulk contribution to the evolution operator (\ref{evolQ1}) now reads
\begin{equation}
\mathbb{W}_{L,\text{bulk}}(\lambda)=-\sum_j\left((F_{j+1}-F_j)(F^+_{j+1}-F_j^+)+(F_{j+1}-F_j)^2F_j^+ F_{j+1}^+\right)
\end{equation}
We represent $\ee^{{\mathbb W}_L(s')t}$ by means of a path-integral~\cite{tailleurkurchanlecomte-2,doi} involving coherent states related to the operators $F$ and $F^+$, which we shall denote by $\phi(\tau)$ and $\bar{\phi}(\tau)$. This leads to an action
\begin{equation}\label{actionbulk}
S_{L,\text{bulk}}[\bar{\phi},\phi]=\int_0^t\dd t\left[\sum_{j=1}^L\bar{\phi}_j\p_t\phi_j+\sum_{j=1}^{L-1}\left[(\phi_{j+1}-\phi_j)(\bar{\phi}_{j+1}-\bar{\phi}_j)+(\phi_{j+1}-\phi_j)^2\bar{\phi}_j\bar{\phi}_{j+1}\right]\right]
\end{equation}
while the boundary terms are given by
\begin{equation}\label{actionbords}\begin{split}
S_{L,\text{boundary}}[\bar{\phi},\phi]=&-\int_0^t\dd t \left[\alpha\bar{\phi}_1-(\alpha+\gamma)\bar{\phi}_1\phi_1\right]\\
&+\ee^{s'}\int_0^t\dd t\left[ \left(  (\ee^{-s'}\beta+\delta)\phi_L-\delta           \right) \left(-\ee^{-s'}+((\ee^{-s'}-1)\phi_L+1)\bar{\phi}_L+1              \right) \right]
\end{split}\end{equation}

Since we are interested in the large-time behavior, we shall proceed with a saddle point approximation at fixed $t$ but as $L\to\infty$, keeping the system size $L$ fixed (this is possible due to our saddle-point equations being stationary). We use the notation $\nabla_j \phi=\phi_{j+1}-\phi_j$. The saddle point equation obtained by differenting $S_L$ with respect to $\bar{\phi}_{j}$ reads
\begin{equation}
(\nabla_j\bar\phi+2\nabla_j\phi \bar{\phi}_j \bar{\phi}_{j+1})-(\nabla_{j-1}\bar\phi+2\nabla_{j-1}\phi \bar{\phi}_{j-1} \bar{\phi}_{j})=0
\end{equation}
and thus there exists $K_1$ such that
\begin{equation}\label{s1}
\nabla_j \phi=\frac{-\nabla_j \bar{\phi}+2K_1}{2\bar{\phi}_j}
\end{equation}
Writing the variational equation with respect to $\phi_j$ and using (\ref{s1}), we obtain
\begin{equation}
\frac{\bar{\phi}_{j+1}+\bar{\phi}_{j-1}}{\bar{\phi}_{j+1}\bar{\phi}_j^2\bar{\phi}_{j-1}}(4K_1^2-\bar{\phi}_j^2+\bar{\phi}_{j+1}\bar{\phi}_{j-1})=0
\end{equation}
which we multiply by
\begin{equation}
\frac{\bar{\phi}_{j+1}-\bar{\phi}_{j-1}}{\bar{\phi}_{j+1}+\bar{\phi}_{j-1}}\bar{\phi}_j
\end{equation}
so that
\begin{equation}
\left[\frac{4K_1^2-(\nabla_j\bar{\phi})^2}{\bar{\phi}_{j+1}\bar{\phi}_j}\right]-\left[\frac{4K_1^2-(\nabla_{j-1}\bar{\phi})^2}{\bar{\phi}_{j}\bar{\phi}_{j-1}}\right]=0
\end{equation}
which leads to the existence of another constant $K_2$ such that
\begin{equation}\label{s2}
(\nabla_j\bar{\phi}_j)^2=4K_1^2+K_2\bar{\phi}_j\bar{\phi}_{j+1}
\end{equation}
We thus obtain that when evaluated at the saddle, $S[\bar{\phi},\phi]=-t\frac{L-1}{2}K_2$. Besides, it is possible to solve the bulk saddle point equations (\ref{s1}) and (\ref{s2}):
\begin{equation}\label{s3}
2\leq j\leq L-1,\;\;\bar{\phi}_j=-A\sinh  [(j-1)B+C],\; \phi_j=E+\frac{1}{2A}\tanh\frac{(j-1)B+C}{2}
\end{equation}
where $A$, $B$ and $C$ are related to $K_1$ and $K_2$ by $K_1=-\frac{A}{2}\sinh B$, $K_2=4\sinh^2 \frac B2$. 
At this stage we write the saddle point equations corresponding to the fields located at the boundaries $j=1$ and $j=L$. At $j=1$ this leads to
\begin{equation}
0=\bar{\phi}_2-\bar{\phi}_1+2(\phi_2-\phi_1)\bar{\phi}_2 \bar{\phi}_1-\alpha\bar{\phi}_1-\gamma\bar{\phi}_1
\end{equation} 
and
\begin{equation}
0={\phi}_2-{\phi}_1-(\phi_2-\phi_1)^2\bar{\phi}_2+\alpha(1-\phi_1)-\gamma \phi_1 
\end{equation}
This immediately sets the constant $E$ appearing in (\ref{s3}) to $E=\frac{\alpha}{\alpha+\gamma}=\rho_0$, and further imposes that $a=\frac{1}{\alpha+\gamma}=\frac{\sinh C}{\sinh B}$. Due to the latter relation between $B$ and $C$, only two unknowns $A$ and $B$ remain to be determined. This is done by writing the two saddle point equations at $j=L$ and by substituting the solution (\ref{s3}). The additional constraints on $A$ and $B$ (or $C$) are
\begin{equation}
A^2=\frac{(z-1)[z(\rho_1-1)-\rho_1]}{A[(z-1)\rho_0+1][z\rho_0(\rho_1-1)-\rho_0\rho_1+\rho_1]},\;\;z=\ee^{-s'}
\end{equation}
and
\begin{equation}
\sinh [(L-1) B+C+\eps]+\frac{b}{a}\sinh C=0
\end{equation}
where $\sinh^2\frac\eps 2=\omega$, $\rho_1=\frac{\delta}{\beta+\delta}$ is also the density at site $L$ and where  the variable $\omega$ is exactly that defined in (\ref{def-omega-SSEP}) with $s'$ instead of $\lambda$. Finally, we eliminate $C$ to obtain $B$ as the solution to
\begin{equation}\label{s4}
\sinh^2\left[(L-1)B+\eps\right]=\big(a^2+b^2+2a b\cosh\left[(L-1)B+\eps\right]\big)\sinh^2 B
\end{equation}
Equation (\ref{s4}) can be solved in powers of $1/L$: to leading order, $B$ and $C$ are $\mathcal O (1/L)$, while $\eps$ is $\mathcal O (1)$, and thus one has $B=\frac 1L\eps=\frac 2L\operatorname{arcsinh}\sqrt{\omega}$. To the next order one has 
\begin{equation}\begin{split}
\psi_L(s')=&\frac{1}{2a}(-1+\sqrt{1+a^2\sinh^2 B})+\frac{1}{2b}(-1+\sqrt{1+b^2\sinh^2 B})
\\
&+(L-1)\sinh^2\frac B2\\
\simeq & \frac{\mu(s')}{L}-\frac{a+b-1}{L^2}\mu(s')+{\mathcal O}(L^{-3})
\end{split}
\end{equation}
This proves the result announced in (\ref{SSEP-lat}).

\subsection{Kipnis-Marchioro-Presutti model}
For the KMP process, one writes a Langevin equation for $\eps_i=\frac 12 x_i^2$ based on (\ref{defLangevinKMP}). Using the It\^o discretization scheme, this leads to
\begin{equation}
\frac{\dd\eps_i}{\dd t}=j_i-j_{i+1}
\end{equation}
where the local energy current is $j_{i+1}=\eps_i-\eps_{i+1}+2\sqrt{\eps_i\eps_{i+1}}\eta_{i,i+1}$ ($1\leq i\leq L-2$), and $j_1=\gamma_1 T_1-2\gamma_1\eps_1+2\sqrt{\gamma_1 T_1}\xi_1$, $j_{L+1}=-\gamma_L T_L+2\gamma_L\eps_L+2\sqrt{\gamma_L T_L}\xi_L$. Using the Janssen-De Dominicis formalism, one is again led to
\begin{equation}
\langle\ee^{-s Q}\rangle=\int{\mathcal D}\bar{\eps}_j{\mathcal D}\eps_j\ee^{-S[\bar{\eps}_j,\eps_j]}
\end{equation}
where the action has the expression
\begin{equation}\begin{split}
S=&\int\dd t\sum_{j=1}^L\bar{\eps}_j\p_t \eps_j+\int\dd t\sum_{j=1}^{L-1}\left[(\bar{\eps}_{j+1}-\bar{\eps}_{j}-s)(\eps_{j+1}-\eps_j)-2\eps_j\eps_{j+1}(\bar{\eps}_{j+1}-\bar{\eps}_{j}-s)^2\right]\\
&+2\gamma_1\int\dd t\left[-T_1(\bar{\eps}_1-s)((\bar{\eps}_1-s)\eps_1+1/2)+(\bar{\eps}_1-s)\eps_1\right]\\
&+2\gamma_L\int\dd t\left[-T_L(\bar{\eps}_L+s)((\bar{\eps}_L+s)\eps_L+1/2)+(\bar{\eps}_L+s)\eps_L\right]
\end{split}\end{equation}
With the change $\bar{\eps}_j^\prime=\bar{\eps}_j-s j$, and dropping the primes, the action becomes
\begin{equation}\begin{split}
S=&\int\dd t\sum_{j=1}^L\bar{\eps}_j\p_t \eps_j+\int\dd t\sum_{j=1}^{L-1}\left[(\bar{\eps}_{j+1}-\bar{\eps}_{j})(\eps_{j+1}-\eps_j)-2\eps_j\eps_{j+1}(\bar{\eps}_{j+1}-\bar{\eps}_{j})^2\right]\\
&+2\gamma_1\int\dd t\left[-T_1\bar{\eps}_1(\bar{\eps}_1\eps_1+1/2)+\bar{\eps}_1\eps_1\right]\\
&+2\gamma_L\int\dd t\left[-T_L(\bar{\eps}_L+s(L+1))((\bar{\eps}_L+s(L+1))\eps_L+1/2)+(\bar{\eps}_L+s(L+1))\eps_L\right]
\end{split}\end{equation}
which shows that the $\langle\ee^{-s Q}\rangle=\langle\ee^{-s(L+1)Q_L}\rangle$, where $Q_L$ is the time-integrated current flowing between site $L$ and the right thermal bath. We shall denote by $s'=(L+1)s$. An additional change of fields, which leaves the bulk part invariant (see~\cite{tailleurkurchanlecomte-2}), allows to further simplify the boundary terms: we set $\eps'=\frac{2\eps}{1+2\bar{\eps}\eps}$ and $\bar{\eps}'=\frac 12 \bar{\eps}(1+2\bar{\eps}\eps)$, and we obtain (dropping the primes)
\begin{equation}\label{action-discrete-KMP}\begin{split}
S=&\int\dd t\sum_{j=1}^L\bar{\eps}_j\p_t \eps_j+\int\dd t\sum_{j=1}^{L-1}\left[(\bar{\eps}_{j+1}-\bar{\eps}_{j})(\eps_{j+1}-\eps_j)-2\eps_j\eps_{j+1}(\bar{\eps}_{j+1}-\bar{\eps}_{j})^2\right]\\
&+2\gamma_1\int\dd t\left[-T_1\eps_1+\bar{\eps}_1\eps_1\right]\\
&-\gamma_L\int\dd t\left[(s'+2\eps_L(1+s'\bar{\eps}_L))(T_L+\bar{\eps}_L(s'T_L-1)) \right]
\end{split}\end{equation}
We differentiate $S$ given in (\ref{action-discrete-KMP}) with respect to $\bar{\eps}_j$:
\begin{equation}
  \Big[ \nabla_j   \eps - 4 \nabla_j     \bar\eps \: \eps_j    \eps_{j+1}\Big] -
  \Big[ \nabla_{j-1}\eps - 4 \nabla_{j-1} \bar\eps \: \eps_{j-1}\eps_{j}  \Big] =0
\end{equation}
where we used the notation $\nabla_j X = X_{j+1}-X_j$. One thus has a constant 
$K_1$ such that
\begin{equation}
  \label{eq:DF_microsaddleKMP}
   \nabla_j  \bar\eps = \frac {K_1+\nabla_j \eps}{4 \eps_j \eps_{j+1}}
\end{equation}
Differentiating now (\ref{action-discrete-KMP}) with respect to $\eps_j$ one has
\begin{equation}
  \nabla_j\bar\eps-\nabla_{j-1}\bar\eps
  +2(\nabla_j\bar\eps)^2\eps_{j+1} +2(\nabla_{j-1}\bar\eps)^2\eps_{j-1} =0
\end{equation}
and
substituting (\ref{eq:DF_microsaddleKMP}) to get an equation on the $\eps_j$'s only, one obtains
\begin{equation}\label{trucsioux}
  \frac{\eps_{j+1}+\eps_{j-1}}{ \eps_{j+1} \eps_j^2 \eps_{j-1}}
    \big(K_1^2-\eps_j^2 + \eps_{j+1} \eps_{j-1}\big)
 =0
\end{equation}
The trick is to multiply (\ref{trucsioux}) by
\begin{equation}
  \frac{\eps_{j+1}-\eps_{j-1}}{ \eps_{j+1}+ \eps_{j-1}}\eps_j
\end{equation}
which leads to
\begin{equation}
 \bigg[ \frac{K_1^2-\big(\nabla_j \eps\big)^2}{\eps_{j+1} \eps_{j}}\bigg] - 
 \bigg[ \frac{K_1^2-\big(\nabla_{j-1} \eps\big)^2}{\eps_{j} \eps_{j-1}} \bigg] =0
\end{equation}
and thus there exists a constant $K_2$ such that
\begin{equation}
  \label{eq:DFh_microsaddleKMP}
  \big(\nabla_j \eps\big)^2 = 
   K_1^2+4K_2 \eps_j \eps_{j+1}
\end{equation}
We  substitute (\ref{eq:DF_microsaddleKMP})
into the bulk part of the action (\ref{action-discrete-KMP}), and we arrive at
\begin{equation}\label{eq:Wbulk_saddle}
 -\frac 1t S_{\text{bulk}}
  =
  \sum_{1 \leq j\leq L-1}
   \frac{K_1^2-\big(\nabla_j \eps\big)^2}{8\eps_{j+1} \eps_{j}}
  =
  -\frac {L-1}2 \,K_2
\end{equation}
We differentiate the action with respect to the fields at the boundaries $\bar\eps_1$ , $\eps_1$: 
\begin{align}
  -(\eps_2-\eps_1)+4(\bar\eps_2-\bar\eps_1)\eps_1\eps_2+
 2\gamma_1\eps_1
 &=0\\
  -(\bar\eps_2-\bar\eps_1)-2(\bar\eps_2-\bar\eps_1)^2\eps_2+
 2\gamma_1(\bar\eps_1-T_1)
 &=0
\end{align}
Differentiating with respect to $\bar\eps_L$ and $\eps_L$ one gets
\begin{align}\label{eq:dSdeBdiscr}
  (\eps_L-\eps_{L-1})-4(\bar\eps_L-\bar\eps_{L-1})\eps_L\eps_{L-1}+
    \gamma_L\Big[2\eps_L+(1-4T_L\eps_L+4\eps_L\bar\eps_L)\lambda-
    T_L(1+4\eps_L\bar\eps_L)s^{\prime 2}\Big]
 &=0\\
  (\bar\eps_L-\bar\eps_{L-1})-2(\bar\eps_{L}-\bar\eps_{L-1})^2\eps_{L-1}
 -2\gamma_L(1+s'\bar\eps_L)(T_L+\bar\eps_L(\lambda T_L-1))
 &=0
\end{align}

We now proceed with solving the microscopic
equations (\ref{eq:DF_microsaddleKMP}) and (\ref{eq:DFh_microsaddleKMP}). We search for a solution in the form
\begin{align} \label{eq:microsaddle_FFhKMP}
  \eps_j = A \sinh \big[2\big( (j-1)B+C\big)\big]
\quad,\qquad
  \bar\eps_j = E + \frac{1}{4A}\tanh \big[(j-1)B+C\big]
\end{align}
where $A$, $B$, $C$ and $E$ are four constants to be determined by
the four saddle point equations at the boundaries.
We first note that, quite remarkably, (\ref{eq:microsaddle_FFhKMP}) is an
exact solution of the microscopic bulk saddle point
equations (\ref{eq:DF_microsaddleKMP}) and (\ref{eq:DFh_microsaddleKMP}), on condition that
\begin{equation}
  K_1 = -A \sinh (2B)
\qquad;\qquad
  K_2 = \sinh^2 B
\end{equation}
One checks that the saddle equations at site $1$ are solved by
\begin{equation}\label{eq:shCshB}
  E = T_1 
\qquad ;\qquad
  \frac12 \frac{\sinh 2B}{\sinh 2C} = \gamma_1
\end{equation}
Eliminating $\gamma_L$ between the saddle equations at site $L$ yields
\begin{equation}
 A^2=\frac{s'\,  (T_L s' -1)}{16 (T_1 s' +1) (T_L+T_1 (T_L s' -1))}
\end{equation}
Substituting this result into (\ref{eq:dSdeBdiscr}), one gets
\begin{equation}
  \sinh\Big[2\big((L-1)B+C+\eps\big)\Big] +\frac{\gamma_1}{\gamma_L}\sinh 2C=0
\end{equation}
where $\eps$ is such that
\begin{equation}
  \sinh^2 \eps=\omega
 \label{eq:defepsilonKMP}
\end{equation}
and $\omega$ is given by
\begin{equation}
  \omega=\lambda \big(T_1-T_L-\lambda T_1T_L\big)
\end{equation}
in accordance with (\ref{def-omega-KMP}). One can now eliminate $C$ using (\ref{eq:shCshB}) and this gives an equation involving $B$ only
\begin{equation}
  \label{eq:eq_B_microKMP}
  \sinh^2 \Big[2(L-1)B  + 2\eps\Big] = 
  \frac{\gamma_1^{-2}+\gamma_L^{-2}+2(\gamma_1\gamma_L)^{-1}\cosh\Big[2(L-1)B  + 2\varepsilon\Big]}{4} \:\sinh^2 2B
\end{equation}
The large deviation function is given by the value of $\mathbb
W_{Q_L}$ at saddle. Combining the bulk
contribution (\ref{eq:Wbulk_saddle}) together with the boundary terms
read from (\ref{action-discrete-KMP}), one obtains
\begin{align}
  \psi_{Q_L}(\lambda) &=
    \frac 12 \gamma_1
   \ -\  \frac 12 \sqrt{\gamma_1^2+\sinh^2 2B} \ + \
        \frac 12 \gamma_L
   \ -\  \frac 12 \sqrt{\gamma_L^2+\sinh^2 2B}
  \nonumber\\ &\quad  
  - \frac{L-1}2 \sinh^2 B
\end{align}
where $B$ is solution of (\ref{eq:eq_B_microKMP}).

Though the expressions are cumbersome, one can still solve the
equation (\ref{eq:eq_B_microKMP}) perturbatively in powers of $L$, by writing $B$ in the form $B=B_0/L+B_1/L^2+\ldots$.  To lowest order, $B$ is of order $1/L$ and
$\eps$ of order $1$: in (\ref{eq:eq_B_microKMP}), the right-hand term is negligible and one obtains $B_0=-\eps=-\operatorname{arcsinh}\sqrt{\omega}$, which yields
the macroscopic fluctuation theory result $\mu(\lambda)=-\frac 12 (\operatorname{arcsinh}\sqrt{\omega})^2$ found previously in (\ref{muomega},\ref{def-omega-KMP}), as expected. To the next order, one gets
\begin{equation}
  \psi_{Q_L}(\lambda) = 
 \frac 1 {L} {\mu}( \lambda )
 \ + \ \frac 1{L^2} \left[
   \left(1-\frac{1}{2\gamma_1}-\frac{1}{2\gamma_L}\right){\mu}(\lambda) 
 \right]
 \ + \
 \mathcal O(L^{-3})
\end{equation}
which matches the announced result (\ref{KMP-lat}).

\end{document}